# Anomalous superconducting diode effect in a polar superconductor

**Robert Kealhofer[1], Hanbyeol Jeong[1], Arman Rashidi[1], Leon Balents[2], and Susanne Stemmer[1,a)]**

[1] Materials Department, University of California, Santa Barbara, California 93106-5050, USA

[2] Kavli Institute for Theoretical Physics, University of California, Santa Barbara, California 93106, USA

[a)] Corresponding author.  Email: stemmer@mrl.ucsb.edu

**Abstract**

A superconductor with broken time reversal and inversion symmetry may exhibit nonreciprocal charge transport, including a nonreciprocal critical current, also known as superconducting diode effect. We report an intrinsic superconducting diode effect in a polar strontium titanate film. Differential resistance measurements reveal a superconducting state whose depairing current is polarity dependent. There is, however, no measurable deviation from Ohmic behavior, implying that this state does not arise from a bulk magnetochiral anisotropy. In the entire measurement range, the only deviation from linearity in the differential resistance is on the edge of the superconducting transition at high magnetic fields, likely due to the motion of flux vortices. Furthermore, the magnitude of the effect is preserved even when the in-plane magnetic field is oriented parallel to the current, indicating that this effect truly does not originate from a bulk magnetochiral anisotropy.

Ohm's law, describing charge transport in metals, is fundamentally "reciprocal," meaning the resistance does not depend on the current's polarity [1]. Interfaces may create non-ohmic devices with non-reciprocal charge transport, such as the well-known current-voltage characteristics of a *p-n* junction or a Schottky barrier [2-4]. Recently, nonreciprocal transport that may be intrinsically possible in noncentrosymmetric crystalline solids in a magnetic field, but in the absence of an engineered interface, has generated significant interest, and many experimental investigations have observed signatures of this nonreciprocal transport in a variety of test structures [5-10]. This nonreciprocal transport is usually identified with the magnetochiral anisotropy (MCA) [5,11]:

$$R = R_0(1 + \gamma \mu_0 H I), \tag{1}$$

where $\gamma$ is the MCA coefficient, whose microscopic origin lies in the strength of spin-orbit coupling and inversion symmetry breaking [11-13], $R$ is the measured current-dependent resistance, $R_0$ is the linear resistance, $I$ is the applied current, and $\mu_0 H$ is the magnetic induction due to the applied magnetic field. A variety of investigators have identified large MCA coefficients in diverse materials systems, including polar semiconductors [8], monolayer transition metal dichalcogenides [5,6] and superconducting systems [14].

The conditions which allow this current-dependent resistance to be expressed - broken inversion and time-reversal symmetry - also permit the observation of a superconducting diode effect in which the critical current's magnitude is polarity dependent. Theoretical explanations have either proposed that the diode effect is interfacial in origin, namely that an asymmetry in the top and bottom surfaces of a superconducting film or flake gives rise to nonuniform screening currents [15,16], from which the diode effect emerges, or that the diode effect is an intrinsic, bulk phenomenon due to the formation of a helical superconducting state with finite-momentum Cooper pairing [17-19]. In both of these cases, time-reversal symmetry is broken by a magnetic field $\mathbf{H}$ applied perpendicular to the superconductor's polar axis, $\mathbf{c}$. In the bulk case, the strength of the diode effect is expected to be linear in the magnetochiral anisotropy, as the MCA is linear in the spin-orbit strength [18,20]. This effect has been observed in a variety of engineered superconducting systems [21,22]. Theoretical analyses, extending the Ginzburg-Landau theory, have generally agreed on a predicted phenomenology for the bulk effect, with an enhancement of the diode effect at moderate to high magnetic fields (around the Pauli limit) with a nonmonotonic temperature dependence [17,19]. Furthermore, as it has been proposed that polar, strongly spin-orbit coupled superconductors may host unconventional, mixed-parity superconductivity that may also have a nontrivial topology [23,24], that these systems would be expected to exhibit large MCA coefficients and therefore large superconducting diode effects has stimulated a broad search for these presumably interrelated phenomena.

In this Letter, we study nonreciprocal transport and the superconducting diode effect in a strained, doped film of $SrTiO_3$, a strongly spin-orbit coupled, polar superconductor [25,26]. We measure the differential resistance at the fundamental harmonic under varying current bias and in an in-plane magnetic field, which allows us to assess the current-voltage characteristic at each point in the magnetic field-temperature ($H$-$T$) plane, while concurrently recording the polarity dependence of the critical current. The phenomenology our measurements reveal differs from that of previous reports in two important ways: in the superconducting state, the magnitude of the superconducting diode effect is maximized at low magnetic fields (compared to either the critical field or the Pauli limit), and in the normal state, the nonreciprocal transport is largely absent until developing at high fields on the edge of the superconducting transition. We also measure the largest "diode effect" toward the lowest temperatures (the diode effect is suppressed by $T_c/2$), which is at odds with the phenomenology seen in [21] and predicted in [17-19], where the magnitude of the diode effect is suppressed at low temperatures. We argue that the absence of the diode effect at higher fields is due to a crossover in the physics determining the critical current: at low fields, geometric factors lead to a critical current (and an accompanying asymmetry) determined by the depairing energy scale, which is connected directly to the nature of the superconducting state, while at higher fields, vortices form, and their depinning (determined by extrinsic factors) determines the critical current. As normal-state nonreciprocal transport only emerges at the highest fields, we argue it is not reflective of a bulk magnetochiral anisotropy as would be expected in a polar system. Moreover, we observe a diode effect of similar magnitude when the in-plane magnetic field is oriented parallel to the current, inconsistent with either a bulk MCA or a purely interfacial origin for the superconducting diode effect.

The device studied is a Hall bar patterned on an epitaxially strained, 70-nm-thick Sm:SrTiO$_3$ film grown on (001) (LaAlO$_3$)$_{0.3}$(Sr$_2$AlTaO$_6$)$_{0.7}$ (LSAT), as described in detail elsewhere [25,27,28] and in the Supplement [29]. The carrier density of the film was $6\times10^{19}$ cm$^{-3}$. The strain gives rise to ferroelectric order at approximately 100 K with a polar axis (**c**) that is oriented out of the film plane [25,26,30]. The $T_c$ is approximately 530 mK [29]. In a dilution refrigerator with a nominal base temperature of 12 mK, the sample is oriented with the magnetic field applied parallel to the film surface and either parallel or perpendicular to the direction of applied current, respectively. The differential resistance is measured while the DC current bias is swept from zero through the superconducting transition, in both directions, and then the magnetic field is stepped out from 0 T. The negative field polarity is measured after warming above $T_c$ to eliminate spurious signals arising from trapped magnetic flux [14]. We repeated this measurement at several temperatures and identified the value of the critical current; a selection of these data is presented in Fig. 1.

The data in Fig. 1 present a phenomenology of a superconducting diode effect that is inconsistent with the theory that has successfully described previous results [17-19,22]. Note that the superconducting diode effect appears most strongly near zero field, not at the Pauli limit, which is approximately 1 T in this film. The field-boundary of the region exhibiting the diode effect is temperature-dependent but in all cases is less than a fifth of the zero-temperature critical field. The superconducting diode effect can also be identified with a region of overall enhancement of the critical current to peak values to approximately five or six times those of the adjacent (in field) region. The diode effect disappears by 250 mK, somewhat below $T_c/2$. This phenomenology is quantitatively similar even when the in-plane magnetic field is applied parallel to the current (lower row of Fig. 1).

We argue that the asymmetry seen in the critical current, concomitant with the low-field region of enhanced critical current, reflects an intrinsic property of the superconducting state, the depairing energy scale, while the critical current at higher fields reflects the physics of vortex motion. Central to this argument is the question of which factors determine the value of the critical current of a thin superconducting film in a parallel magnetic field, which we now discuss.

In type-II superconducting thin films, due to geometrical effects, when the field lies parallel to the film surface, the critical current density ($J_c$) in the Meissner state (below $H_{c_1}$) is enhanced to values approaching the theoretical depairing current density ($J_d$), the current for which the superfluid velocity is greater than the Cooper pair binding energy [31-33]:

$$J_c = \left(1 - \frac{3\sqrt{3}\pi\xi d}{2\phi_0}H\right)J_d,$$  (2)

where $\xi$ is the coherence length, $d$ the film thickness, and $\phi_0$ the flux quantum. From Eq. (2), we see that $J_c$ is suppressed linearly in increasing magnetic fields, observed in our data to a similar degree as in Stejic et al. (ref. [31]), where this expression is derived. Above $H_{c_1}$, vortices enter the superconducting bulk, and the critical current is determined by balancing the Lorentz force with an extrinsic pinning force, whose strength is determined by microstructural details and may include collective terms due to vortex-vortex interaction [32]. (The thermodynamics governing the formation of vortices in a type-II superconductor is sensitive to sample geometry. In a thin film, $H_{c_1}$ is thickness-dependent [31].) These results also contradict an interfacial origin, in which the diode effect emerges due to an asymmetric distribution of screening currents, as in that case the diode effect would not be suppressed above $H_{c_1}$ (see also the discussion further below).

We next address in more detail the transition between the regimes and the lack of conventional nonreciprocal transport or MCA. Investigations in the literature have identified second-harmonic resistance signals they attribute to an enhanced MCA coefficient in the superconducting state as well as in the normal state. Our group has previously reported large second-harmonic signals in the superconducting state in thick, partially strain-relaxed, uncapped SrTiO₃ films, while not detecting such signals in coherently strained samples similar to the one which is the object of our present study [34]. Here, by measuring differential resistance rather than the second harmonic resistance, we are able to interrogate the existence of nonreciprocal behavior in the normal state, above the critical current, but in the regions of the *H-T* phase diagram that exhibit a superconducting diode effect (i.e., a nonreciprocal critical current).

Representative differential resistance traces recorded at 50 mK and various in-plane magnetic fields are presented in Fig. 2. These traces individually reflect the systematic evolution of the transition width [Fig. 3(a)], critical current magnitude [Fig. 3(b)], and deviation from linearity [Fig. 3(c)] with applied magnetic field, which we now discuss.

For the smallest fields [Figs. 2(a) and 2(d)], there is a substantial difference in critical current between the positive and negative current polarities. The transition to the normal state is extremely sharp, and the differential resistance on either side of the transition is visually flat, implying ohmic behavior. The sharpness of the resistive transition is consistent with our earlier claim that the critical current in this low-field region reflects the depairing critical current, rather than being dominated by physics related to vortices. Comparison with Eq. (1) offers that there may be no visible nonlinearity in the differential resistance inside the superconducting state as the "linear resistance", $R_0$, is zero. In the normal state, however, at bias currents exceeding the critical current, there is also no deviation from ohmic behavior, despite $R_0$ not being zero. Crucially, this indicates the absence of a significant bulk MCA in the regime where we observe the superconducting diode effect.

As the field is increased [Figs. 2(b) and 2(e)], the difference in critical current for opposite bias current polarities disappears, while the transition begins to broaden; the differential resistance remains flat. In this field region, vortices penetrate the superconducting bulk. Their depinning determines the critical current and results in a finite transition width, that, when normalized to the magnitude of the critical current, approaches values of order 1, consistent with what we observe in Fig. 3(a) [35]. That there is no nonlinearity in the differential resistance above or below the transition, save necessarily at the transition itself, indicates the continued absence of a bulk MCA in this state.

Finally, at high fields [Figs. 2(c) and 2(f)], close to the upper critical field, there is substantial curvature to the differential resistance, along with significant width in the transition to the normal state. This region may be identified as part of the broad transition to the normal state. Moreover, pronounced nonlinearity appears in the differential resistance, both below and above the transition to the normal state. Comparison with Eq. (1) would imply that this is due to a bulk MCA, measurable due to the finite resistance of the "superconducting" state on the edge of the transition at the upper critical field [5]. However, this nonlinearity may equally be attributed to vortex motion (e.g. flow, creep, etc.) [36,37]. We note also that this nonlinearity in the differential resistance would contribute to a measurable second harmonic signal, and so caution is indicated in assigning the MCA coefficient to the measured ratio of the second harmonic resistance to the "linear" resistance in this field regime. Comparison with our previous work [34] makes this point succinctly: As in that study, the fully strained film here under study would show no second harmonic resistance, implying that there may be no link between these two species of nonreciprocal transport.

That we observe a diode effect of similar magnitude when the in-plane magnetic field is oriented parallel to the current places two constraints on the origin of the effect. First, it supports our assertion that a bulk MCA is not the origin. That is, when a spin-orbit coupled superconductor with a polar axis $\mathbf{c}$ is placed in a magnetic field, the MCA may produce a diode effect that can be described as

$$J_c \propto 1 + (\boldsymbol{c} \times \boldsymbol{H}) \cdot \hat{\boldsymbol{J}}, \tag{3}$$

with $\hat{\boldsymbol{J}}$ a unit vector parallel to the current [20]. This relation would imply no diode effect when $\mathbf{H}$ and $\mathbf{J}$ are parallel, in contrast to the data we report here. Second, a diode effect may naturally emerge in any heterostructure with asymmetric top and bottom surfaces and a perpendicular in-plane magnetic field [15,16]. However, this mechanism precludes the existence of a diode effect when the field is oriented parallel to the current. To see this, consider the film coordinates depicted in Fig. S1(b) [29]. The mirror plane perpendicular to the film surface and the current, $\sigma_y$, acts to exchange the direction of current, $\boldsymbol{J} \rightarrow \sigma_y \boldsymbol{J} = -\boldsymbol{J}$, while the parallel magnetic field is unaffected as it transforms as a pseudovector, $\boldsymbol{H} \rightarrow \det(\sigma_y) \sigma_y \boldsymbol{H} = \boldsymbol{H}$. Any configuration which preserves $\sigma_y$, as the asymmetric interface mechanism does, cannot yield a diode effect when field and current are oriented parallel to each other. Subtle field misalignments also do not satisfactorily explain our data, although they have been relevant to other investigations [38] (for a discussion, including additional measurements in out-of-plane fields, see [29]).

Taken together, our results reveal a superconducting diode effect, intrinsic to our polar superconducting film, and separate from other nonreciprocal transport effects. This effect is suppressed above the lower critical field, as in this state the critical current is not determined by an intrinsic energy scale but by the details of vortex depinning. Unexpectedly, this superconducting diode effect exists in the absence of bulk magnetochiral anisotropy, and is surprisingly insensitive to the direction of the in-plane magnetic field, in contrast with theoretical predictions and other recent experiments, and calls into question the relationship between magnetochiral anisotropy, the superconducting diode effect, and unconventional superconductivity. To motivate future work, what we report here is largely consistent with proposals for exotic superconducting states in polar superconductors, such as topological superconducting states, or mixed-parity states, in which an anisotropy in the superfluid density is reflected in the critical current [39]. For this case, a careful theoretical consideration of the symmetries, suggested in ref. [40], should be carried out, as our data only permit speculation on this point.

**Acknowledgements**

The authors wish to thank Dr. Nikola Maksimovic and Prof. Liang Fu for sharing fruitful discussions. This work was supported by the U.S. Department of Energy (Award No. DE-SC0020305). A portion of this work was performed in the UCSB Nanofabrication Facility, an open access laboratory. The dilution refrigerator in which many of the measurements were performed was funded through the Major Research Instrumentation program of the US National Science Foundation under Award No. DMR 1531389. The work made use of the MRL Shared Experimental Facilities, which are supported by the MRSEC Program of the U.S. NSF under Award No. DMR 1720256.

## Figure Captions

**Figure 1:** Superconducting diode effect. (a) The critical current as a function of in-plane magnetic field applied perpendicular to the current at 50 mK. (b) The difference between the magnitudes of the positive and negative critical current and is formed by adding the traces in panel (a) (i.e. $\Delta I_c = I_c^+ + I_c^-$). (c) The magnitude of the difference in critical current at various temperatures. A marked asymmetry in the critical current is seen to develop below 250 mK at low fields, coinciding with a low-field region of enhanced critical current. (d) The critical current as a function of in-plane magnetic field applied parallel to the current at 50 mK. (e) The critical current asymmetry in the parallel field orientation qualitatively behaves as in the perpendicular orientation above, and (f) evolves similarly with increasing temperature.

**Figure 2**: Differential resistance traces and transition widths at 50 mK. (a-c) Representative differential resistance traces reveal the field evolution of the transition width and non-Ohmic resistive component as the perpendicular in-plane magnetic field is varied. The lower traces (dashed) are measured at negative current bias polarity, while the upper curves (solid) are offset by 200 $\Omega$ and measured at positive current bias polarity. At the lowest fields, the transition is sharp and the resistance is ohmic. Above the lower critical field, a finite transition width emerges, and near the upper critical field pronounced nonlinearity develops in the differential resistance. (d-f) Similar behavior is seen in representative traces collected with the in-plane magnetic field oriented parallel to the current bias.

**Figure 3:** (a) The transition width normalized to the value of the critical current versus applied magnetic field. The color-coded regions correspond to magnetic fields below the lower critical field (green), the mixed state (pink), and the transition region near the upper critical field (blue); compare with Fig. 2. Above the lower critical field, the transition width increases substantially.

(b) Critical current versus magnetic field. The solid curve is the critical current with positive bias current polarity, while the dashed curve is at negative bias current polarity. (c) The deviation from linearity in the normal state (solid) and superconducting state (dashed) versus applied magnetic field.

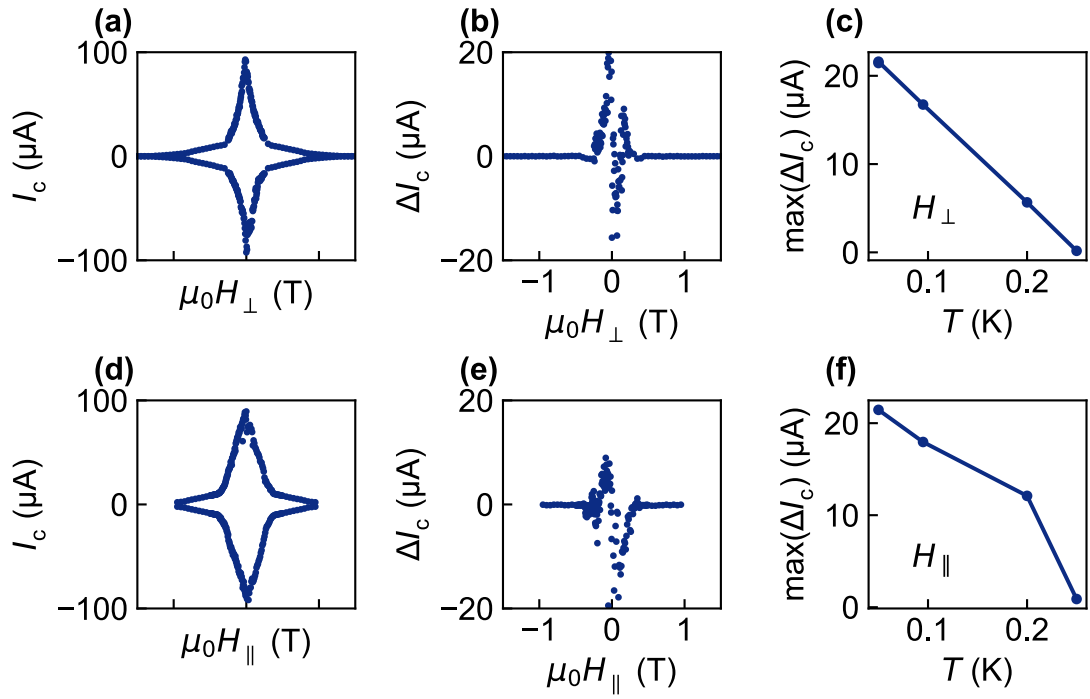

Figure 2

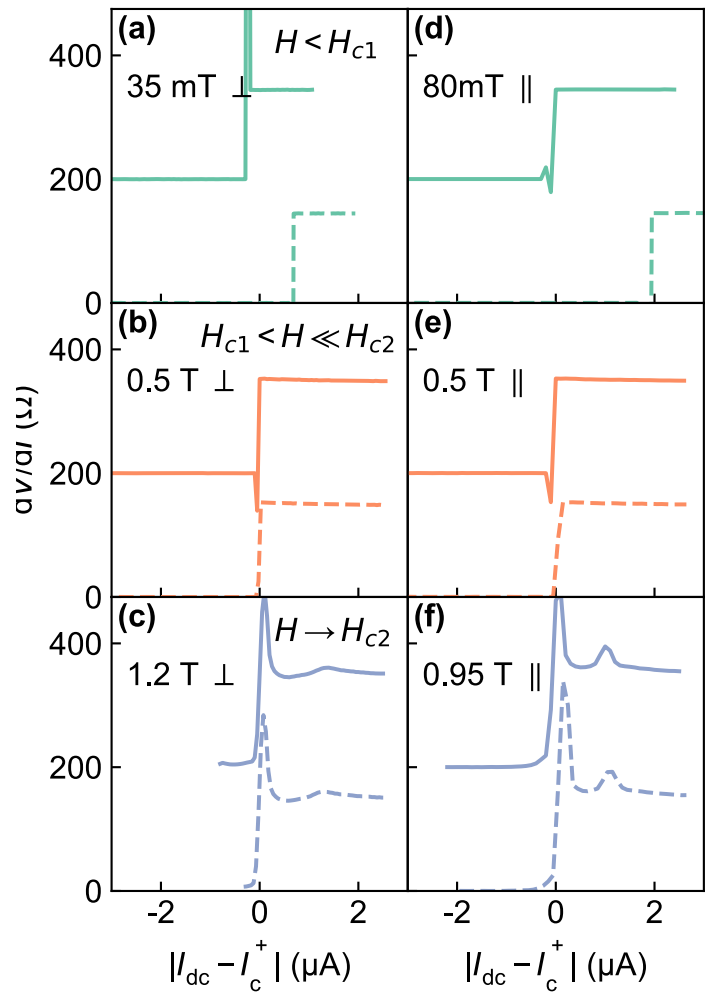

**Figure 3**

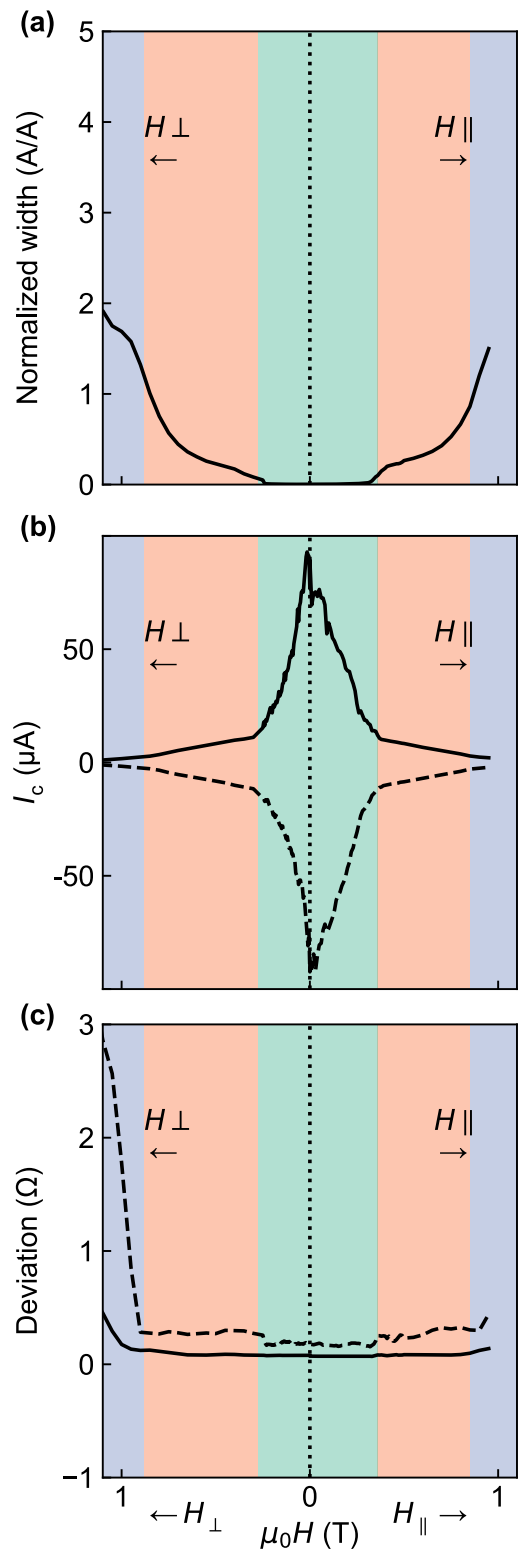

# Anomalous superconducting diode effect in a polar superconductor

**Robert Kealhofer[1], Hanbyeol Jeong[1], Arman Rashidi[1], Leon Balents[2], and Susanne Stemmer[1,a)**

[1] Materials Department, University of California, Santa Barbara, California 93106-5050, USA

[2] Kavli Institute for Theoretical Physics, University of California, Santa Barbara, California 93106, USA

## Sample Structure and Structural Characterization

The device studied is a Hall bar patterned using photolithography on an epitaxially strained Sm:SrTiO$_3$ film, shown schematically in Figs. S1(a) and S1(b). The film was grown by hybrid molecular beam epitaxy on (001) (LaAlO$_3$)$_{0.3}$(Sr$_2$AlTaO$_6$)$_{0.7}$ (LSAT) substrates, as described in detail elsewhere [1,2]. The carrier density was determined from Hall effect measurements to be 6×19 cm$^{-3}$. To prevent surface depletion of the carriers, the film is capped with a layer of EuTiO$_3$, which preserves the carrier density necessary to observe superconductivity [3]. The thickness of the film, determined from a combination of growth calibrations, RHEED oscillations [1], x-ray thickness oscillations is 70 nm [3].

High-resolution X-ray diffraction measurements were performed using a Philips Panalytical X'Pert thin-film diffractometer with Cu Kα radiation. The position of the LSAT substrate 002 reflection is 46.944 degrees, and the position of the STO film's 002 reflection is 46.132 degrees. The lattice mismatch with LSAT results in an in-plane compressive strain of 0.9%, consistent with the position of the 002 reflection in x-ray diffraction scans, shown in Fig. S1(c). Previous work on similar films indicates that the strain gives rise to ferroelectric order at approximately 100 K [2,4,5]. The polar distortion induced by the compressive strain supplied by the substrate is oriented out of the film plane, along the polar structure's tetragonal **c** axis depicted in Fig. S1(b). The superconducting $T_c$ is measured to be approximately 530 mK [Fig. S1(d)]. This critical temperature is the temperature at which the resistance reaches 50% of its normal state value.

## Fabrication and measurement

Hundred-micron "Hall bar" devices were fabricated using optical photolithography, Ar ion milling for mesa isolation, and electron beam deposition of Ti/Au Ohmic contacts. The Hall bar device is 100 microns wide with an aspect ratio of 1 and 6 voltage taps. Low frequency AC current-biased resistance and differential resistance measurements at temperatures between 12 mK and 1 K were performed in an Oxford Instruments Triton dilution refrigerator with low-pass filters installed at the mixing chamber plate [6]. Voltages were measured using a Stanford Research Systems SR860 lock-in amplifier; current bias was supplied by a SRS CS580 current source. AC current bias was 100 nA while DC bias was swept using the DC level set on the reference voltage output from the SR860.

In the cryostat, a dilution refrigerator with a nominal base temperature of 12 mK, the sample is oriented with the magnetic field applied parallel to the film surface and either parallel or perpendicular to the direction of applied current, respectively [Fig. S1(b)]. The differential resistance is measured while the DC current bias is swept from zero through the superconducting transition, in both directions, and then the magnetic field is stepped out from 0 T. The negative field polarity is measured after warming above the critical temperature and cooling in zero field to eliminate the possibility of a spurious signal arising from trapped magnetic flux [7]. Critical currents reported are determined from outgoing (increasing in magnitude from zero) sweeps of the DC current bias. To avoid hysteretic effects from the intermediate superconducting state, measurements at positive and negative magnetic fields were performed after separate zero-field cooldowns. We repeated this measurement at several temperatures and identified the value of the critical current; a selection of these data is presented as Fig. 1 in the main text.

**Data processing**

The X-ray data presented are raw data. The resistance vs. temperature trace showing the critical temperature is downsampled by a factor of 200 to minimize figure size.

The critical current is determined by thresholding the value of the measured differential resistance at 66% of the normal state value.

The differential resistance traces are processed by averaging the recorded resistance at each DC current value (typically 4-5 values). The transition width is determined by fitting a step function to the critical current, subtracting it, and determining the full width of the resulting lineshape within half an ohm of the normal state or zero resistance value.

The deviation from linearity is the standard deviation of the trace values away from the transition region. This quantity is calculated as follows. First, a step function is constructed from the zero-bias trace value and the high-bias trace value. This step function is then subtracted from the trace so that in the "ideal" case the transition appears as just a single point that deviates from a background of zero differential resistance above and below the transition. Then, the points in the transition width (calculated above) are discarded. The standard deviation of the remaining points is then calculated. The plotted values represent the average for both positive and negative current bias at each field value.

**Sample alignment error and low-field artifacts**

An experimental challenge is that these data are collected in a superconducting solenoid designed to achieve 14 T magnetic fields, with no provision for sample rotation or vector fields. Field alignment and field accuracy are then two issues that bear addressing, especially given recent work by Hou et al. proposing that out-of-plane field components must be sensitively zeroed to convincingly observe in-plane superconducting diode effects.

First, we address the question of field accuracy. Without a field sensor at the position of the sample, we rely on the current sourced by the magnet power supply to determine the field in the 14 T solenoid. This introduces two errors in the determination of the field, which are significant: contributions from the magnet's remanent field, and contributions from leakage across the persistent mode switch (as the magnet is operated in "driven" mode). Both of these errors are independent and of the order 10-20 mT. The error due to field remanence is not a constant offset and is history dependent. We have made no serious attempt to compensate for these errors as we do not judge them to qualitatively alter our findings; in turn, we have not attempted quantitative fits or analyses at low fields and welcome future work addressing this deficiency. We have not performed any analysis that hinges on the absolute value of "zero" magnetic field (notably, our data processing does not involve antisymmetrizing in field).

Second, we address the issue of out-of-plane field components. Our sample is mounted using vacuum grease on a machined copper block which is in turn screwed to a machined bracket. The surface tension of the grease ensures co-planarity of the film surface to the machined block at room temperature. The out-of-plane angular misalignment between these various machined surfaces is unlikely to exceed a fraction of a degree.

The mount is designed to place the sample at the field center. The finite thickness of the sample (half a millimeter) means that the field felt by the device on the substrate surface may not be entirely in-plane (our solenoid is not infinite in length). The question is now whether an out-of-plane field produces a large diode effect to which we may attribute our in-plane results.

In Figure S2, we present the critical current and its asymmetry as in Fig. 1 in the main text for the field oriented out-of-plane, along with the scaled transition width as in Fig. 2 in the main text. It is clear that 1) the geometric enhancement of the critical current, allowing us to probe the depairing critical current at low fields, is absent, 2) there is a markedly different suppression of critical current with out-of-plane magnetic field, 3) the diode effect is absent at all fields, and 4) the rapid suppression of the critical current in the magnet's remanent field (specifically at "zero" applied field) limits our observed critical current to a value that is 30% of the value we observe with the field oriented in-plane. We add that these results, as well as the behavior we describe in the main text, depart significantly from those of Hou et al.

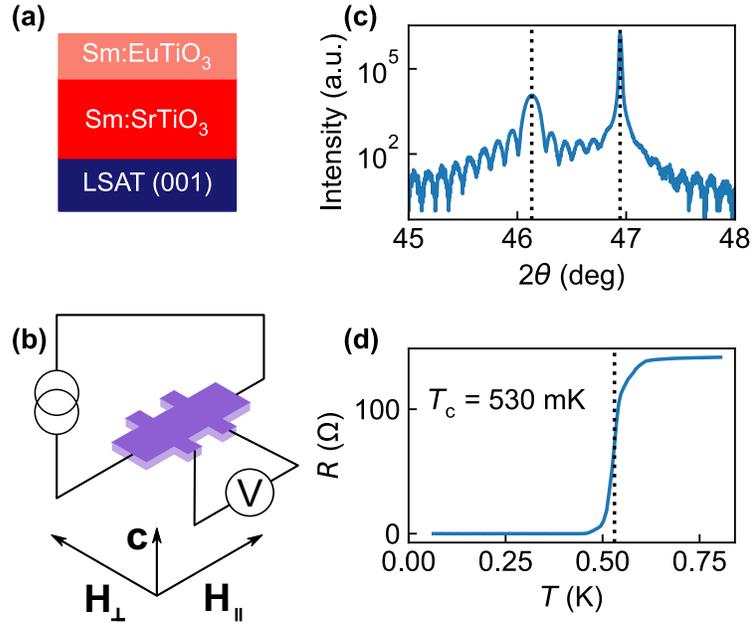

**Figure S1:** Sample schematics and characterization. (a) Schematic of the heterostructure studied in this letter. A 70-nm coherently strained SrTiO₃ film is grown on a (001) LSAT substrate, with a thin EuTiO₃ capping layer to prevent surface depletion. (b) A schematic of the Hall bar device used in these measurements and the orientations of the magnetic field applied in this study. (c) X-ray diffraction 2θ-ω scan showing the 002 film and substrate peaks. The film peak appears at 46.3 degrees, consistent with the expected out-of-plane lattice parameter for a coherently strained film on LSAT. The film thickness from the Laue fringes on this peak agrees with the 70 nm thickness determined from RHEED oscillations during growth. (d) The superconducting transition, measured on warmup from base temperature. The transition temperature is reported here as the temperature at which the resistance is half its normal state value.

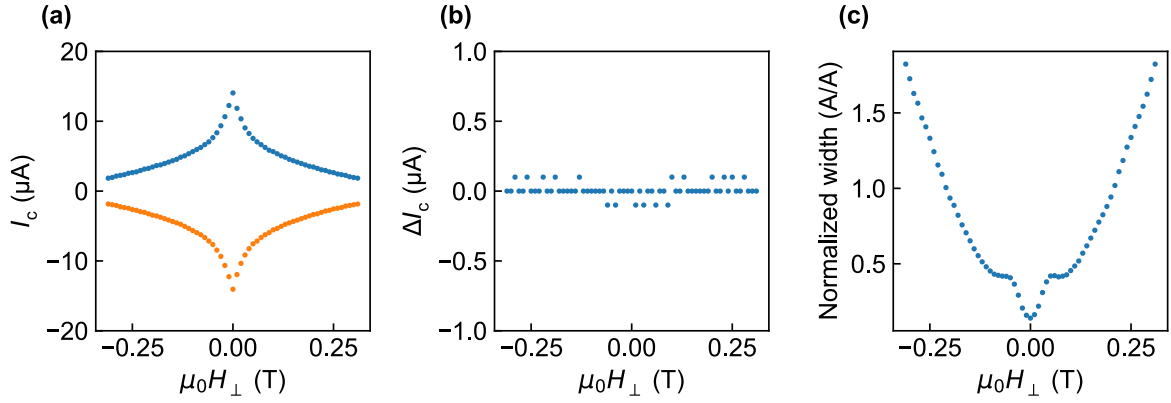

**Figure S2:** Critical current measurements with the field oriented out-of-plane. (a) Rapid reduction in critical current with applied magnetic field. Remanent field of approximately 10-20 mT is enough to suppress critical current below "zero field" values in parallel field orientation. (b) No diode effect is visible at any field value. (c) The transition width is generally broad, as the critical current is not determined by the depairing energy scale at any field value.

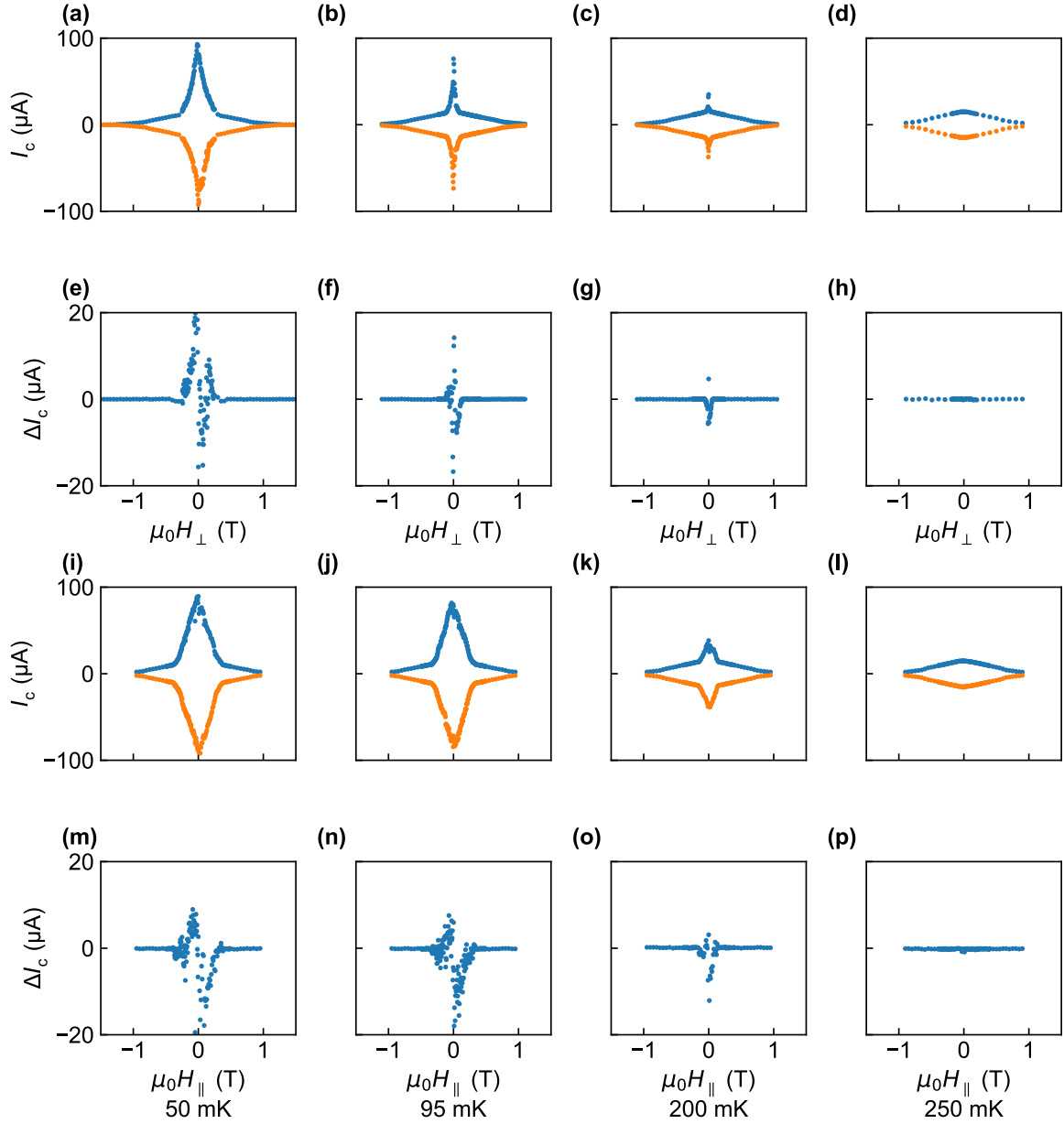

**Figure S3:** Critical current and difference in critical current as in Figs. 1(a) and (d), collected at various temperatures. Note that the disappearance of the low-field region of enhanced critical current at 250 mK corresponds to the disappearance of the superconducting diode effect.

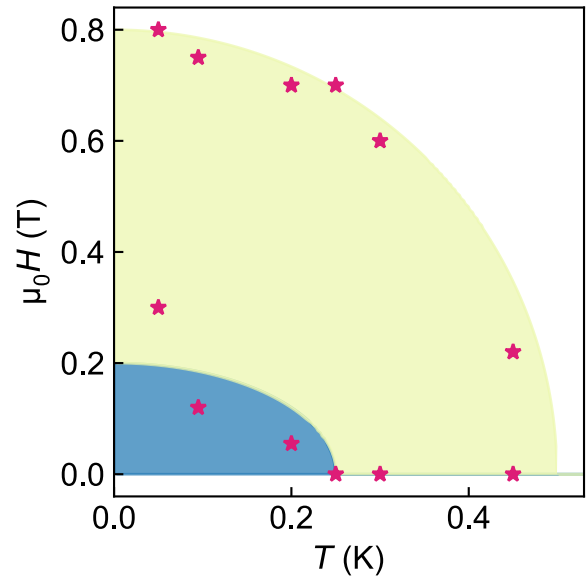

**Figure S4:** The field- and temperature-dependence of the superconducting diode effect is diagrammed. The lower boundary (corresponding to the schematic blue region) is the region which shows a superconducting diode effect; the upper boundary (corresponding to the schematic yellow region) is the onset of the upper critical field, above which nonlinearity develops in the differential resistance. Note that these colored regions are intended schematically, not as a fit to the data.